\documentclass{sig-alternate-10pt}

\PassOptionsToPackage{hyphens}{url}  
\usepackage[hyphens]{url}
\usepackage{booktabs}  
\usepackage{multirow}
\usepackage{xspace}
\usepackage{changepage} 
\usepackage{flushend} 
\usepackage{pifont}
\usepackage{graphicx}
\usepackage{subfigure}
\usepackage{authblk}

\usepackage{tikz}
\usepackage[inline]{enumitem}
\usepackage{enumitem}

\newcommand{\squishlist}{
 \begin{list}{$\bullet$}
  { \setlength{\itemsep}{0pt}
     \setlength{\parsep}{3pt}
     \setlength{\topsep}{3pt}
     \setlength{\partopsep}{0pt}
     \setlength{\leftmargin}{1.5em}
     \setlength{\labelwidth}{1em}
     \setlength{\labelsep}{0.5em} } }

\newcommand{\squishend}{
  \end{list}  }

\newcommand{\hide}[1]{}

  {\end{list}}

\newcounter{ecount}
  {\end{list}}

\makeatletter \newcommand{\listoftodos}
{\section*{Todo List} \@starttoc{tdo}}
\newcommand{\l@todo}
{\@dottedtocline{1}{0em}{2.3em}} \makeatother

\makeatletter
\let\@copyrightspace\relax
\makeatother

\newtoggle{techreport}
\toggletrue{techreport}

\begin{document}
 
\renewcommand{\paragraph}[1]{\vspace{0.1in}\noindent\textbf{#1}}
\newcommand{\paratight}[1]{\vspace{0.05in}\noindent\textbf{#1}}
\newcommand{\parait}[1]{\vspace{0.1in}\noindent\textit{#1}}
\newcommand{\paraittight}[1]{\vspace{0.05in}\noindent\textit{#1}}

\title{Revisiting Comparative Performance of DNS Resolvers in the IPv6 and ECS Era\titlenote{This work was supported by NSF through grant CNS-1647145.}}
\numberofauthors{1}
\author{
{\large Rami Al-Dalky and Michael Rabinovich} \\
\affaddr{Case Western Reserve University}
\email{\{rami.al-dalky,michael.rabinovich\}@case.edu}
}


\maketitle

\begin{abstract}
This paper revisits the issue of the performance of DNS resolution services available to Internet users.  While several prior studies addressed this important issue, significant developments, namely, the IPv6 finally getting traction and the adoption of the ECS extension to DNS by major DNS resolution services, warrant a reassessment under these new realities.
We find that DNS resolution services differ drastically -- by an order of magnitude in some locations -- in their query response time. 
We also find established resolvers (Google DNS and OpenDNS) to lag far behind relative newcomers (Cloudflair and Quad9) in terms of DNS latency, and trace the cause to drastically lower cache hit rates, which we further trace to less cache sharing within the resolver platform.  
 In addition, we find that public resolvers have largely closed  the gap with ISP resolvers in the quality of CDNs' client-to-edge-server mappings as measured by latency.
 Finally, in most locations, we observe IPv6 penalty in the latency of client-to-CDN-edge-server mappings produced by the resolvers.   Moreover, this penalty, while often significant, still does not rise above typical thresholds employed by the Happy Eyeballs algorithm for preferring IPv4 communication. 
 Thus, dual-stacked clients in these locations may experience suboptimal performance.  
 \end{abstract}

\section{Introduction}
\label{sec:intro}

DNS is a core component of the Internet machinery that not just maps human-readable hostnames into IP addresses but also plays a vital role in traffic engineering.  In particular, CDNs, which, according to Cisco \cite{Cisco_VNI}, deliver over half of Web traffic to consumers, 
commonly use DNS to assign an end-user to the appropriate edge server.  The performance of the DNS system, and especially the quality of DNS-provided mapping of a user to an edge server when the user accesses CDN-delivered content, plays a direct impact on user's Web experience.  This paper considers the performance of the DNS system from the end-user perspective, both in terms of its latency in answering queries and in terms of the quality of CDN user mappings.  Several studies considered these important issues (e.g., \cite{ager2010comparing,huang2011public,hours2016study}) and in particular documented lower user mapping quality produced by  public resolvers.  However, the emergence of the EDNS-Client-Subnet (ECS) DNS extension \cite{rfc7871}  to help public DNS resolvers provide high-quality CDN mappings for their users, as well IPv6 finally getting traction \cite{CAZ+14}, warrant a reassessment of these issues under these new realities. 

This paper compares the performance experienced by end-users when using four popular IP public resolvers in terms of DNS resolution time and the quality of CDNs' client-to-edge-server mappings, referred henceforth as "client mappings". Furthermore, all public resolvers and CDNs we consider are dual-stack, that is, capable of communicating with their clients over both IPv4 and IPv6.  Thus, by recruiting vantage points that are also dual stack, we are able to directly assess any impact the IP version may have on these performance aspects.  

Our main contributions include the following novel findings.  

\squishlist
\item
We find that DNS resolution services differ drastically – by an order of magnitude in some locations – in their query response time. In particular, the latency of established resolvers (Google DNS and OpenDNS) in our measurements far exceeds the latency of relative newcomers (Cloudflare and Quad9). We present strong evidence that a major cause of these higher latencies lies in less cache sharing within the resolvers platforms. Prior studies that considered public resolvers performance \cite{ager2010comparing,huang2011public,hours2016study} focused on comparing them to ISP resolvers rather than to each other.  A notable exception is a non-peer-reviewed NANOG presentation \cite{DNS_resolvers_NANOG18}, which also observed some of the  differences but did not uncover the causes behind them.  We contrast our findings with those in \cite{DNS_resolvers_NANOG18} later in this paper.

\item
We assess the impact of the IP version choice on DNS latency of the interaction between DNS clients and their resolvers.  While multiple prior studies compared general performance of IPv4 and IPv6 (e.g., \cite{law2008empirical,nikkhah2011assessing,zhou2008ipv6,wang2005understanding}), 
they only consider communication performance, whereas DNS latency may also be affected by server platforms, potentially engineered and provisioned differently for IPv4 and IPv6.  We find that the DNS latency is generally little affected by the client's choice of IP version to interact with the DNS system.

\item
While IP version has little impact on DNS latency, we do observe IPv6 penalty in the latency of CDNs' client mappings in most of our locations.  Moreover, this penalty, while often substantial, still does not rise above typical thresholds employed by the Happy Eyeballs algorithm for preferring IPv4 communication. Thus, clients in these locations who choose IPv6 for Internet communication or use the Happy Eyeballs algorithm \cite{rfc8305} to dynamically select between IPv4 and IPv6 for TCP communication, may experience suboptimal performance when accessing CDN-accelerated content. 

\item
Finally, we find that public resolvers have all but closed the gap with ISP-provided resolvers in the quality of CDNs' client-to-edge-server mappings as measured by latency documented previously in a number of studies \cite{ager2010comparing,huang2011public,hours2016study}. 
 \squishend
Our measurement datasets are available at \cite{our_data}.


\section{Related Work}
\label{sec:related}

Several studies investigated the impact of using DNS public resolvers on end-users' performance \cite{ager2010comparing,huang2011public,hours2016study}. 
While differing in methodologies, they found that ISP resolvers 
were geographically closer \cite{huang2011public}, 
and  redirected end-users to more proximal CDN edge servers \cite{ager2010comparing,huang2011public,hours2016study} than the public DNS resolvers considered (collectively, Google, OpenDNS, and Level 3). However, these studies were conducted  before ECS was either proposed \cite{ager2010comparing,huang2011public} or adopted by the CDN under study \cite{hours2016study}. None of them consider the impact of IPv6 on DNS behavior.  Further, they focus on comparing public resolvers performance to that of ISP resolvers rather than to each other.

A NANOG presentation \cite{DNS_resolvers_NANOG18} compares the responsiveness and availability of a large number of public resolvers but limits its analysis to only client-resolver interaction.  An important aspect of our study, not addressed in \cite{DNS_resolvers_NANOG18}, is the comparison of quality of CDN client mappings produced by different resolution services. With respect to the latencies of client-resolver interactions, we contrast our findings with \cite{DNS_resolvers_NANOG18} later in the paper but also uncover the root causes behind performance differences among the public resolvers in this regard (Section~\ref{sec:DNS_Latency_Diffs}).


Chen et. al \cite{chen2015end} studied the impact of enabling ECS at Akamai on the quality of client-to-edge-server mapping, especially for the clients using public resolvers. Their results show that enabling ECS has decreased the RTT between these clients and their edge servers by 50\%, and significantly improved other metrics, at the cost of increasing the number of DNS queries from public resolvers to Akamai's authoritative DNS servers by a factor of 8. S\'anchez et. al \cite{sanchez2013dasu} found similarly significant impact of ECS on the quality of client mapping in the EdgeCast CDN for clients using Google Public DNS. Using active measurements from a specially instrumented client application, they observed the reduction in the time to obtain the first byte of content of 20-60\% for clients in North America and Western Europe and 70-90\% for clients in Oceania. At the same time, our study shows that, with sufficient resolver footprint, a public DNS resolver can provide competitive client-to-edge-server mappings without resorting to ECS as we found Cloudflare and, to a less extent, Quad9, achieve this for most regions and CDNs we consider.

Turning to the impact of IPv6 transition, Alzoubi et.al \cite{alzoubi2013performance} studied performance implications of unilateral enabling of IPv6 by Websites. They found no evidence of performance penalty for doing so, although their measurements employed coarse time granularity of 1 second. This finding was largely confirmed by Bajpai et. al  \cite{bajpai2016measuring}.
Probing Alexa top-10K websites from 80 vantage points, the authors found that although most tested websites had higher latency over IPv6, 91\% of these sites had IPv6 latencies within 1 msec of their IPv4 counterparts. Our investigation complements these studies by considering IPv6 impact on the quality of client-to-edge-server mappings, and finds the impact to be much more significant. \vspace{-2mm}
%

\section{Methodology}

\begin{figure}[tb]
    \centering
          \includegraphics[width=0.9\columnwidth]{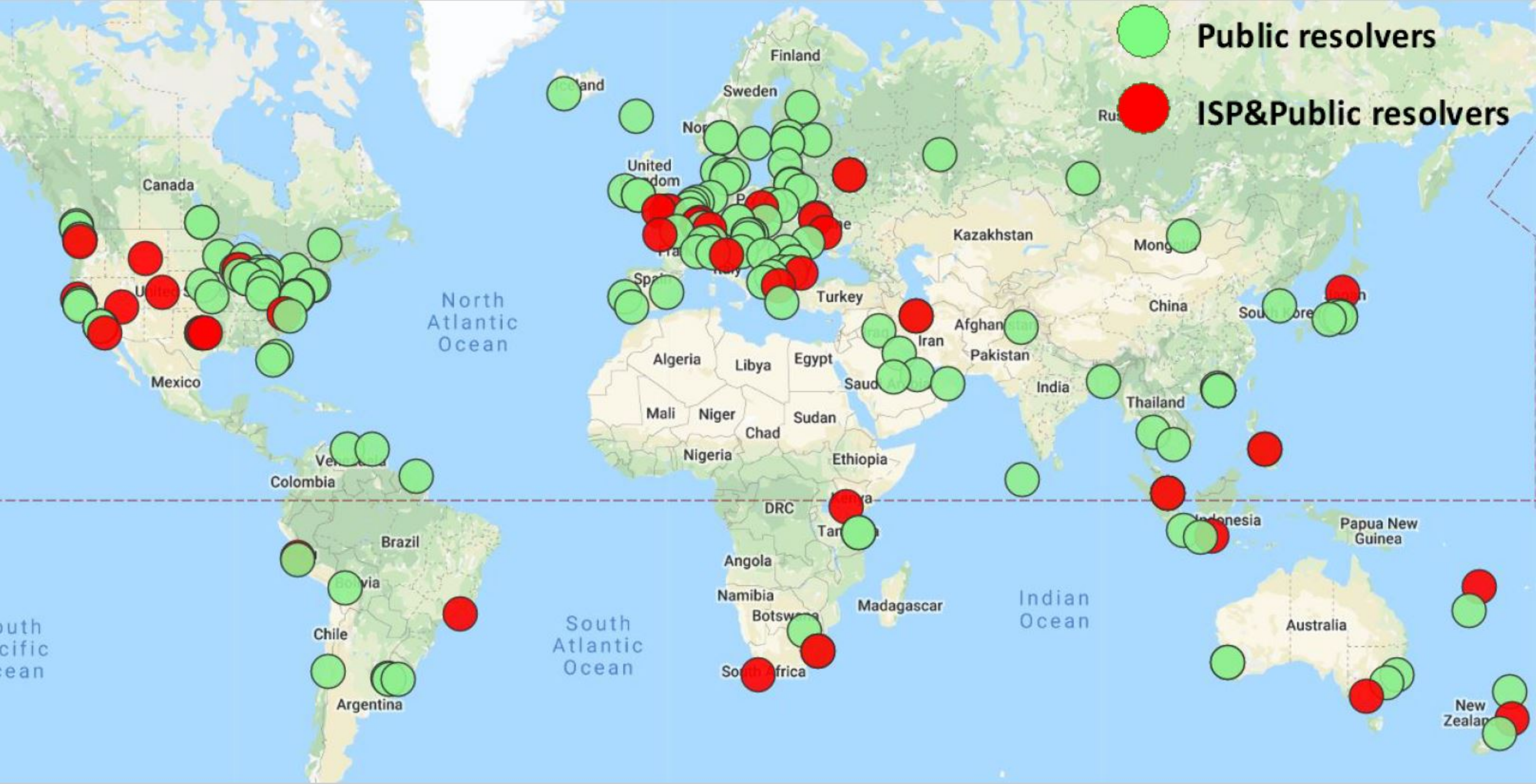}
          \vspace{-1em}
      \caption{Distribution of probes.} \vspace{-2mm}
      \label{fig:probes_dist}
\end{figure}

To conduct this study, we use 200 dual-stack RIPE Atlas probes \cite{RIPE_Atlas}, chosen from the total $~$1600 probes listed as dual-stack by RIPE Atlas based on their stability and diversity of represented autonomous systems and geographic locations.  Using more probes would not appreciably improve general representativeness of our results because of a general strong skew of RIPE Atlas probes towards North American and, especially, European locations, which are already disproportionally represented in our sample.  Four of our RIPE Atlas probes failed to get allocated to our experiment and did not produced any results. Moreover, we excluded another 8 probes as they consistently timeout on some services/protocols. The 188 productive probes are distributed in 74 countries across 188 ASes and 6 regions: 52  in North America (US and Canada), 70 in Europe, 38 in Asia, 11 in Latin America, 7 in Africa, and 10 in Oceania (a region that includes Australia and Pacific islands). 
Figure~\ref{fig:probes_dist} shows the distribution of the probes. 

We use these vantage points to access the first 100 websites from Majestic top-1M list \cite{maj} that (a) support both IPv4 and IPv6 protocols (b) are accelerated by a CDN -- which we determine by examining the CNAME chain of the DNS resolution of a website with "www"  prepended  and (c) support HTTPS such that we can measure the latency between the probes and the assigned edge server\footnote{RIPE Atlas probes don't support HTTP requests but do allow a TLS handshake, thus enabling the TCP 
latency measurement between the probe and the CDN edge server to which the probe is mapped.}.
The CDNs used by these websites include Akamai (65 sites), Cloudfront(17), Google(9), Fastly(8), and Incapsula (1).

\noindent{\bf Results Representativeness:}
Two aspects of our measurements concern the representativeness of our results: our selection of vantage points and the websites used to collect our datasets.  With regard to the vantage points, while we carefully selected our probes to represent diverse geographic regions and ASes, one can't assume they follow the distribution of the user concentrations.  Thus, our results provide point assessments of comparative client experience at our vantage points as they choose different resolvers and IP versions, but we caution against using them for Internet-wide generalizations\footnote{We still note that the number of vantage points in our study compares favorably with prior peer-reviewed studies, including \cite{ager2010comparing}, which used "more than 60 vantage points" from 50 commercial ISPs, \cite{hours2016study}, which employed a single vantage point, and \cite{huang2011public}, which performed its study from the perspective of a single website and its content delivery platform. A NANOG presentation \cite{DNS_resolvers_NANOG18} describes a study that utilized somewhat more IPv4 vantage points (252 vs. 188) but less than a third of IPv6 vantage points (58 vs. 188) than our study.}. 

With regard to using only 100 websites, we emphasize that our goal is not to assess the performance of the websites themselves but to compare the quality of client-to-edge-server mappings a CDN provides when accessed through various DNS resolution services and IP versions.  Since Akamai, with its vast footprint of relatively small points of presence, is known to utilize only subsets of their points of presence for delivery of individual websites, it is useful to probe its mapping behavior through several websites.  But generally, a small number of busy websites per CDN suffices. To verify this, we test that our results are not skewed by the website selection.  We pick 50 websites out of the original 100 by randomly selecting roughly half of each CDN's customers: 32 from Akamai, 8 from Cloudfront, 5 from Google, 4 from Fastly, and 1 from Incapsula.  We then consider what effect, if any, this smaller set of websites would have on our results.
We observe the median DNS latencies to stay within 2 msec from their values with 100 websites, and the median mapping latency within 1 msec for Akamai and within 2msec for non-Akamai CDNs.

\noindent {\bf Measurements:}  We consider the following public resolvers: Google Public DNS, OpenDNS, Quad9 and Cloudflare. 
We further contrast performance of these public resolvers with the ISP-provided resolvers for the probes.  For the latter, we only consider the probes for which we can determine that their default resolvers are provided by their ISP.  We determine this is the case by sending a query to our own domain from each probe and inferring the probe uses an ISP resolver if the two belong to the same autonomous system as determined by Team Cymru \cite{cymru} or share at least a /24 prefix. 
Of all our productive probes, 36 pass this test and we use measurements from these probes when considering performance of ISP-provided resolvers.  These probes represent 23 countries in all 6 regions, and 36 autonomous systems. 

We conduct the following measurements when exploring a website from each probe.  First, we measure DNS query latencies when querying for IPv4 address (A-type query) and IPv6 address (AAAA-type query), and when communicating with the resolver over IPv4 or IPv6 protocol -- resulting in four combinations of query types and protocols used for communication.  
Second, we assess the quality of the obtained CDN mappings by measuring the latency of TCP handshake to the received CDN edge server, using its obtained IPv4 and IPv6 addresses. We conducted our experiment between October 30 and November 4, 2018.

For DNS latency, since we are interested in the responsiveness of the resolution services themselves, we attempt to factor out uncertainties due to unpredictable state of caching at the time of measurement by putting all resolution services on the level playing field with regard to caching.  We thus first send a query to prewarm the revolver cache and, after waiting for 10 sec to ensure this query does precede the measurement despite Atlas's imprecise scheduling (since even unused records have been shown to typically stay in the DNS cache much longer \cite{schomp2013measuring}, the response is likely to remain the cache despite this delay), use the median response time of three subsequent back-to-back queries as the measurement result for analysis. For mapping latency, we perform three downloads of the SSL certificate from the assigned CDN edge server
and take the median handshake RTT (time between the SYN and SYN/ACK segments) for the result.  We refer to this metric as {\em mapping latency} below. We interchangeably call assignments of clients to IPv4 (resp., IPv6) edge servers as IPv4 or A (resp., IPv6 or AAAA) mappings.  To ensure fairness of the analysis, unless a probe produces all results (both DNS and mapping latencies) through every resolver for a specific website, we exclude such (probe,website) pair from any further consideration. Thus, we are able to use 17,573, out of $188 \times 100 = 18.8K$ possible, such pairs.  

\begin{figure}[tb]
    \centering
          \includegraphics[width=0.9\columnwidth]{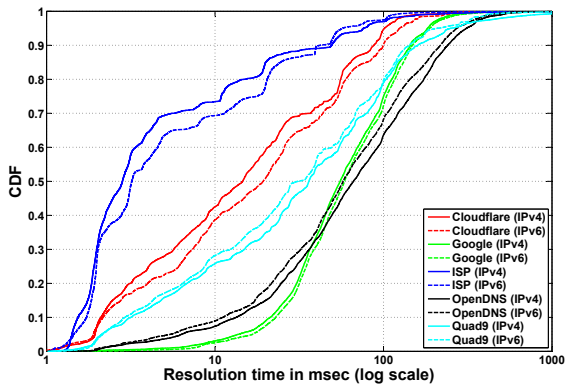}
      \caption{Distribution of DNS latencies for A-type queries over IPv4 and IPv6.} \vspace{-5mm}
      \label{fig:DNS_Latency_v4}
\end{figure}

\section{DNS Response Time}
\label{sec:DNS-latency}


We first consider the overall distribution of DNS response times across all our vantage points, which reflects the aggregate performance trends for various resolution services and protocols, and then present performance differences from individual vantage points perspective.   Figure~\ref{fig:DNS_Latency_v4} plots cumulative distribution of response times for A-type DNS queries conducted over IPv4 and IPv6 protocols using the resolvers under study (the distributions of AAAA query latencies are virtually identical). 
For convenience, Table~\ref{tbl:medianDNSLatencies} lists median latencies for all query types and protocols. 
We make the following observations.


\begin{table*}[tb]
 \begin{center}
 \begin{tabular}{|c|c|c|c|c|c|c|c|c|c|c|}
  \hline
IP   & \multicolumn{2}{c|}{ISP-Provided} & \multicolumn{2}{c|}{Cloudflare} & \multicolumn{2}{c|}{Quad9}  & \multicolumn{2}{c|}{Google} & \multicolumn{2}{c|}{OpenDNS} \\ \cline{2-11}
& A & AAAA & A & AAAA & A & AAAA & A & AAAA & A & AAAA \\ \hline \hline
v4 & 2.82& 2.78&14.24& 14.56 & 36.03 & 34.80 &55.29 & 57.75 & 66.24 &61.96 \\ \hline
v6 &3.21 & 3.19& 19.27& 19.28& 31.06& 31.44& 58.67& 57.84 & 57.51& 55.38 \\ \hline 
 \end{tabular}
  \caption{Median DNS latencies (ms).  The first column shows the IP version  used to interact with a resolver.  Other columns show DNS latency of these interactions for a given resolver and query type.} \vspace{-8mm}
  \label{tbl:medianDNSLatencies}
  \end{center}
\end{table*}

\squishlist
\item
The times to resolve A and AAAA queries via a given resolver over a given IP version are very close.  From Table~\ref{tbl:medianDNSLatencies}, median latencies for both query types are within 5 ms of each other.  
\item The DNS latency is also generally similar for IPv6 and IPv4 interactions. Delving  deeper, Cloudflare responds somewhat slower\footnote{Throughout the paper, whenever we point out a difference in distributions, we verified that the Kolmogorov-Smirnov test rejected the hypothesis that the two samples in question come from the same distribution at significance level of at most 0.1\%, and in fact mostly at much lower significance levels as the p-values are vanishingly small.} over IPv6 than over IPv4, with its median latency 35\% and 32\% higher for A and AAAA queries, respectively, while Quad9 and OpenDNS are actually slightly quicker. We discuss possible reasons for these better latencies in Section~\ref{sec:DNS_Latency_Diffs}.
Google's performance is virtually unaffected, and ISP resolvers are so much quicker than the public resolvers that any impact from the protocol choice is immaterial by comparison.     
\item ISP resolvers respond statistically much faster than  public resolvers, both over IPv4 and IPv6.
\item Among public resolvers, longer established providers (Google and OpenDNS) are significantly slower than relative newcomers (Quad9 and especially Cloudflare).  
\squishend

Comparing our results with the findings in the NANOG presentation (see \cite{DNS_resolvers_NANOG18}, slide 12), both studies find lower latency of Cloudflare than Google and OpenDNS.  However, we find the latencies for all three providers, as well as the latency difference between Cloudflare and the other two providers, significantly higher than in \cite{DNS_resolvers_NANOG18}.  We can attribute our generally greater latencies to the fact that our vantage points represent a variety of networks including residential ones (which tend to have higher last-mile latencies), whereas \cite{DNS_resolvers_NANOG18} used vantage points in their own data centers.  However, this can't explain the greater difference in latencies between the providers, for which we don't have an explanation.  

\begin{figure}[tb]
    \centering
          \includegraphics[width=0.9\columnwidth]{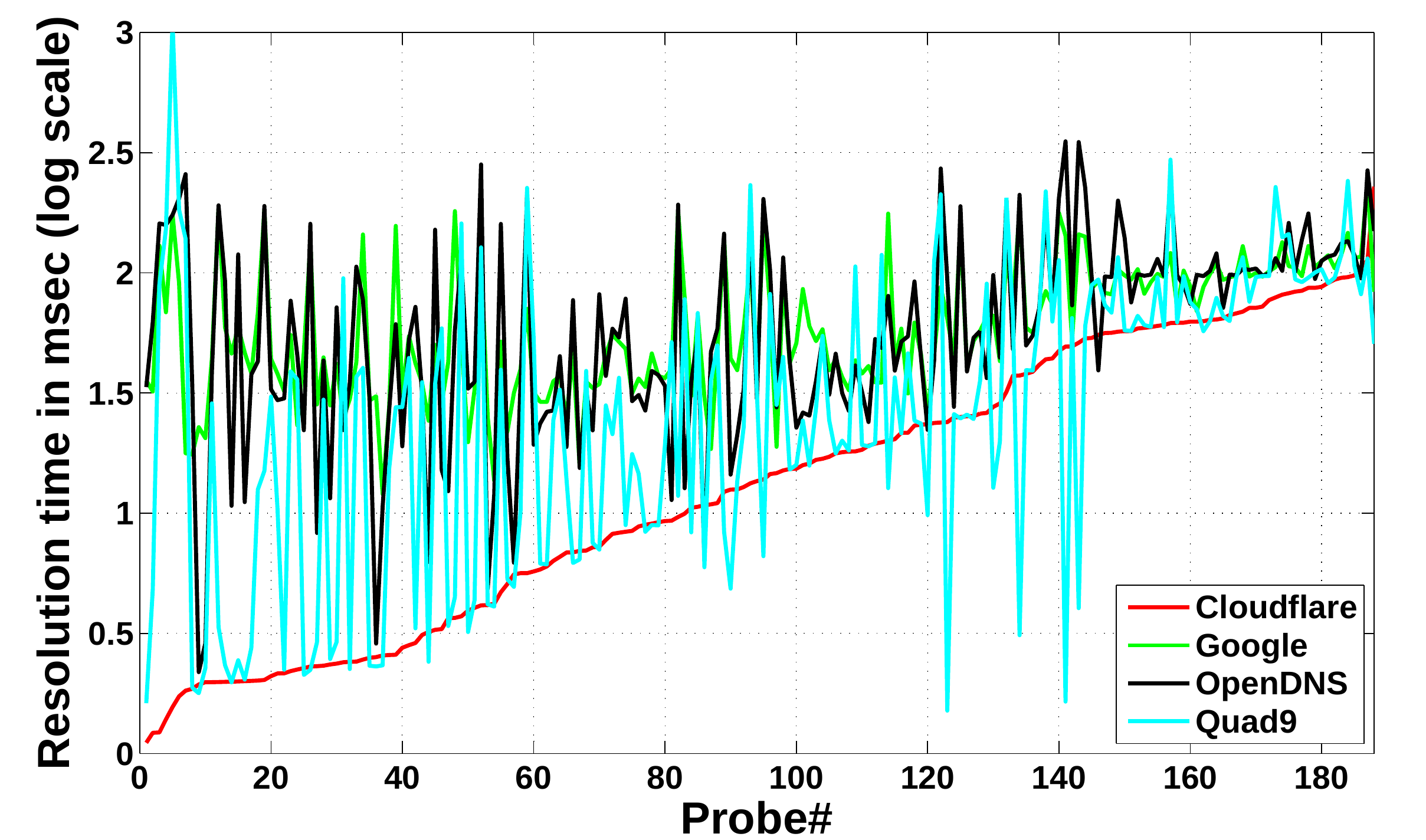}
      \vspace{-1em}
      \caption{Average DNS response times of public resolvers for A-type queries over IPv4 as seen by each vantage point.} \vspace{-5mm}
      \label{fig:DNS_Latency_Per_VP_A_IPv4}
\end{figure}
\vspace{-0.2em}

We now turn from aggregate DNS latency distributions across all vantage points to DNS latency comparison from the perspective of each vantage point.  Figure ~\ref{fig:DNS_Latency_Per_VP_A_IPv4} shows the DNS response times seen by each vantage point when using the public resolution services under study. Each data point is the median response time of a given resolver over the 100 hostname queries used in our study. The vantage points are listed on X-axis in the order of their Cloudflare latencies, and the remaining curves show corresponding latencies these vantage points experience through other resolvers.  We only present the results for A-type queries over IPv4; the other three combinations of query type and the IP version produced similar results.  

The figure shows that the conclusion we drew earlier for aggregate latency distributions across vantage points also holds for individual vantage points: a vast majority of vantage point experience significantly better average DNS latency with Cloudflare and Quad9 than with Google and OpenDNS, and a great majority of vantage points have lower latency with CLoudflare than with Quad9, although for some locations Quad9 holds large advantage.  The figure further shows that for quite a few locations, these latency differences can be dramatic, by an order of magnitude or more, while other locations see only marginal differences.  
Thus, if one wants to maximize their DNS performance, they need to test different DNS providers from their specific location before choosing one to use.  

\subsection{On Causes for DNS Latency Differences}
\label{sec:DNS_Latency_Diffs}
Lagging performance of Google and OpenDNS relative to Cloudflare has been observed previously in a non-peer reviewed article \cite{DZone}, but we would like to understand possible reasons behind this finding.  One factor could be a difference in footprints.  However, ping latencies from our probes to the resolvers' anycast front-ends \footnote{Because we did these measurements later, some of our probes were no longer available; we had 139 probes for IPv4 pings and 130 for IPv6.} paint a mixed picture: while Cloudflare indeed shows lower median latencies (11.01ms for IPv4 and 12.85ms for IPv6) than the other three (24.49ms and 25.41ms for Quad9, 18.55ms and 23.67ms for Google, and 25.13ms and 26.82ms for OpenDNS), the other three resolvers all have latencies in the same ballpark. Thus, at least their front-ends are at a similar distance. 
We uncover another reason for the performance difference: the data below provides strong indication that the difference in the DNS cache miss rates plays a major role.  

To assess the miss rate, we obtain the authoritative TTL of the responses by directly querying the authoritative DNS servers of the CDN services used by our target websites. Then we consider a response to be a cache miss if its TTL is equal to the authoritative except for Google, which  decrements authoritative TTL by 1 sec {\em before} serving the authoritative response on a miss \cite{google_cache}. So for Google we detect a cache miss if the response TTL is  one second less than the authoritative.


Table~\ref{tbl:miss_rates} lists miss rates of the different resolvers for A queries, as well as the median latencies of the queries that hit and missed in the cache.  The results for AAAA queries have the same trends and are not shown.  In this analysis, we include all three queries past the cache-warming query for each website.  


\begin{table}[t]
 \begin{center}
 \begin{tabular}{|c|c|c|c|c|c|c|}
   \hline
          & \multicolumn{2}{c|}{Miss rate } & \multicolumn{2}{c|}{Hit latency }  & \multicolumn{2}{c|}{Miss latency} \\  
Resolver   & \multicolumn{2}{c|}{(\%)} & \multicolumn{2}{c|}{(ms)}  & \multicolumn{2}{c|}{(ms)} \\ \cline{2-7} 
& IPv4 & IPv6 & IPv4 & IPv6 & IPv4 & IPv6 \\ \hline \hline
ISP    &6.26 &7.20 &2.53 &3.05 &31.14 &24.30  \\ \hline
{\small Cloudflare} & 1.76 & 2.26 &14.46&19.16 &24.07&28.64   \\ \hline
Quad9 & 16.67 & 17.22 &34.16 &28.31 &156.32 &68.55 \\ \hline
Google & 72.47& 75.85 & 30.97 &31.50 &64.05 &65.44 \\ \hline
{\small OpenDNS} & 83.11 & 72.94 & 24.08 &26.47 & 77.89 & 76.63 \\ \hline
 \end{tabular}
  \caption{DNS miss rates for A-type queries.} \vspace{-8mm}
  \label{tbl:miss_rates}
  \end{center}
\end{table}

Table~\ref{tbl:miss_rates} provides a clear sources of the lower latencies of Cloudflare and Quad9 over Google and OpenDNS.  First, both Quad9 and, especially, Cloudflare have dramatically lower miss rate. In fact, in the case of Quad9, this lower miss rate compensates for much {\em higher} miss latency, still resulting in lower overall latency in Figure~\ref{fig:DNS_Latency_v4}\footnote{As a side observation, we also note a large difference between Quad9's miss latency over IPv4 vs. IPv6. While speculating on a reason for this finding would require understanding of their platform architecture, this difference explains why Table~\ref{tbl:medianDNSLatencies} shows lower median latency of  Quad9 over IPv6 than IPv4.}.   Second, Cloudflare has significantly lower latencies for both hits and misses than the other public resolvers.  

\iftoggle{techreport}{
\begin{table*}[t]
\footnotesize
 \begin{center}
\scalebox{0.9}{
 \begin{tabular}{|c|c|c|c|c|c|c|c|c|c|c|}
  \hline
        & \multicolumn{2}{c|}{Akamai}  & 
        \multicolumn{2}{c|}{Google} & 
        \multicolumn{2}{c|}{Fastly}  & \multicolumn{2}{c|}{CloudFront}   & \multicolumn{2}{c|}{Incapsula}  \\ \cline{2-11}
Resolver& Hit & Miss & Hit & Miss & 
         Hit & Miss & Hit & Miss &
         Hit & Miss  \\ \hline \hline
Cloudflare & 14.60 & 20.38 & 14.00 & 58.43 & 14.00 & 31.04 & 14.39 & 56.38 & 13.68 & 2775 \\
Quad9      & 34.82 & 148.14 & 30.96 & 176.95 & 33.16 & 103.86 &  34.27 & 228.45 & 35.88 & 124.08 \\
Google     & 30.94 & 68.82 & 31.67 & 39.95 & 30.83 & 51.26 &  30.86 & 66.51 & 30.89 & 49.91 \\
OpenDNS    & 23.88 & 80.33 & 24.12 & 69.80 & 24.03 & 37.26 &  26.40 & 133.31 & 24.65 & 42.98 \\
 \hline   \hline
\end{tabular}
}
  \caption{Median DNS A/IPv4 query latencies for hits and misses (msec).}
  \label{tbl:medianDNSLatencies_A_per_CDN}
  \end{center}
\end{table*}

\begin{table*}[t]
\footnotesize
 \begin{center}
\scalebox{0.9}{
 \begin{tabular}{|c|c|c|c|c|c|c|c|c|c|c|}
  \hline
        & \multicolumn{2}{c|}{Akamai}  & 
        \multicolumn{2}{c|}{Google} & 
        \multicolumn{2}{c|}{Fastly}  & \multicolumn{2}{c|}{CloudFront}   & \multicolumn{2}{c|}{Incapsula}  \\ \cline{2-11}
Resolver& Hit & Miss & Hit & Miss & 
         Hit & Miss & Hit & Miss &
         Hit & Miss  \\ \hline \hline
Cloudflare  & 19.30 & 26.07  & 18.60 & 58.48  & 18.97 & 26.12 &  19.15 & 51.47 & 17.91 & 2748 \\
Quad9      & 29.60 & 60.95 & 26.71 & 89.26 & 28.31 & 55.81 &  28.32 & 129.60 & 26.66 & 69.91 \\
Google      & 31.81 & 70.55 & 27.59 & 43.58 & 30.43 & 51.86 & 30.41 & 67.78 & 36.10 & 44.13 \\
OpenDNS    & 23.64 & 85.02 & 25.50 & 70.59 & 25.23 & 41.25 & 30.12 & 116.37 & 29.59 & 44.31 \\
 \hline   \hline
\end{tabular}
}
  \caption{Median DNS AAAA/IPv6 query latencies for hits and misses (msec).}
  \label{tbl:medianDNSLatencies_AAAA_per_CDN}
  \end{center}
\end{table*}
}{}
While Cloudflare's lower hit latencies probably reflect its wider footprint as mentioned earlier, we were initially puzzled to see Cloudflare {\em miss} latencies to be lower than others' {\em hit} latencies.  
A closer examination, however, revealed that this  phenomenon is only present for hostnames from Akamai's domain\iftoggle{techreport}{(see Tables~\ref{tbl:medianDNSLatencies_A_per_CDN} and ~\ref{tbl:medianDNSLatencies_AAAA_per_CDN} for per-CDN hit and miss latencies of the resolveres under study\footnote{We note the extremely high Cloudflare-to-Incapsula miss latencies; we can only explain it by "bad luck", as there were only a very small number of these missed queries in our measurement -- four A query misses and two AAAA query misses across all the probes.})}, and the overall result also reflects this phenomenon due to the dominance of Akamai-accelerated hostnames in our dataset.  
Akamai uses widely distributed anycasted ADNS \cite{AKAM_DNS} and its own private backbone for communication among its points of presence \cite{AKAM_Backbone2018}.  Combined with the extensive footprint of Cloudflare, this leads to a situation where many vantage points have a nearby Cloudflare resolver, which further has a nearby replica of Akamai ADNS. Either a miss in Cloudflare cache would be resolved from that nearby ADNS directly, or -- even if it needs to travel to a remote central ADNS -- this communication would presumably occur over Akamai's own highly optimized backbone, resulting in a lower latency than in the other public resolvers with smaller footprints. 

What could be a reason for high miss rates at some public resolvers? Given that we prewarm the cache before taking our measurements,
For the same reason, static cache fragmentation among anycast endpoints or ECS are unlikely causes -- although the latter can play a role as discussed later.  Indeed, multiple queries from the same 
probe are likely to arrive at the same anycast endpoint and belong to the same client subnet, hence our cache prewarming should have populated even a statically fragmented cache.  We thus conjecture that the most likely cause for higher miss rates is dynamic server rotation with limited cache sharing, especially that cache fragmentation in public resolvers has been noted previously \cite{moura2018dike}. We verify this conjecture in two steps. 

First, to verify server rotation, we employ a special DNS record provided by Akamai that allows a client to learn its egress resolver and the resolver's ECS behavior.  Specifically, a TXT-type query for whoami.ds.akahelp.net will return the IP address of the egress resolver that communicates with Akamai's authoritative DNS server, along with other information such as the ECS prefix (which we make use of below).  We send 100 queries for the above record from our lab host through each of the four public resolvers we consider, at 1-sec. intervals, and observe 32 egress resolvers from Google, 13 from OpenDNS, 7 from Quad9, and 6 from Cloudflare. Thus, public resolvers do rotate their servers, and Google and OpenDNS more so than Quad9 and Cloudflare, at least from our vantage point\footnote{While theoretically the resolvers could follow different server rotation policies for different clients, we don't believe this is likely.  Unfortunately we can't not verify this by observing server rotation  from our probes because RIPE Atlas rate limiting prevents repeated queries for the same name from the same probe.}. 
\begin{figure}[t]
    \centering       \includegraphics[width=0.9\columnwidth]{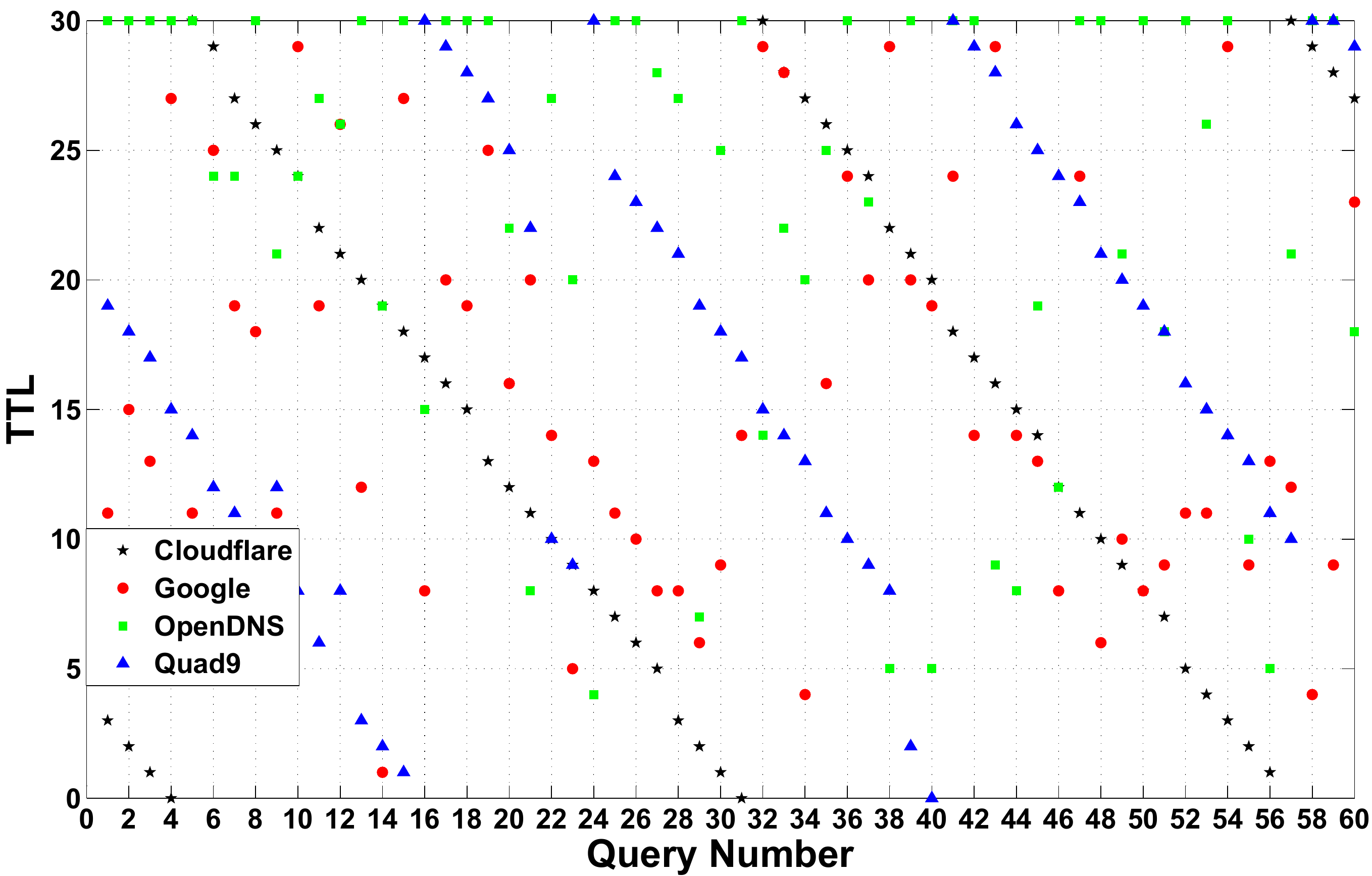}
    \vspace{-1em}
      \caption{Response TTLs of repeated queries for a Fastly-accelerated hostname.} \vspace{-5mm}
      \label{fig:cacheRotation}
\end{figure}

Second, to detect if these rotating servers share their caches, we send repeated queries for the same hostname to a given resolver, 1 second apart. 
As an example, Figure~\ref{fig:cacheRotation} shows a scatter graph of these queries, plotting the query number vs. TTL of the answer, for the hostname served by Fastly (other CDNs show the same behavior).  Queries (even if not consecutive) served from the same cache should fall roughly along a declining straight line\footnote{The line may not be perfectly strait because successive queries arrive at the cache slightly more than 1 sec apart due to network delay,
which may cause the TTL to decrease by 2 sec from one query to the next.}.  
The graph shows perfect cache sharing for Cloudflare: the data points form uninterrupted lines until TTL expires and then start again.  Quad9 mostly uses the same cache but occasionally routes queries though other caches.  However, in the Google and OpenDNS cases, the points spread along a number of descending lines, and there are many isolated points as well.  
We conclude that despite the Google-provided description of its public DNS architecture with a global shared cache \cite{google_dns}, in reality, Google (and OpenDNS) suffers from cache fragmentation.  We verified these results with 100 queries for our own domain, observing 26 egress resolvers for Google and 14 for OpenDNS, vs. 5 for Quad9 and 4 for Cloudflare, and seeing explicit queries at our authoritative DNS server for each cache miss inferred through TTLs.   

This penalty can be exacerbated by the usage of ECS, which may render cached records not usable for some queries. Indeed, Akamai, which accelerates two-third of our websites and thus cotributes a majority of our data points, is known to support ECS with Google and OpenDNS \cite{akamai_ECS_support}, while Quad9 and Cloudflare do not use ECS \cite{quadECS,cloudflareECS}).  
In fact, ECS effects can also explain the lower miss rate exhibited by OpenDNS over IPv6 compared to IPv4.  We query whoami.ds.akahelp.net through Google and OpenDNS over both IPv4 and IPv6 and observe that, while both convey the same /24 prefix for IPv4 clients, their ECS prefixes for IPv6 clients differ: Google conveys /56 client subnet prefix while OpenDNS conveys /48 client subnet prefix. Assuming Akamai commonly returns the ECS scope of the same length (which could be expected given their wide footprint), OpenDNS can reuse the responses for more clients. \vspace{-2mm}


\section{CDN Mapping Quality}
\label{sec:mappings}

\begin{figure*}[tb]
    \centering     
     \begin{subfigure}[Akamai (all latencies)]
         {\includegraphics[width=0.7\columnwidth]{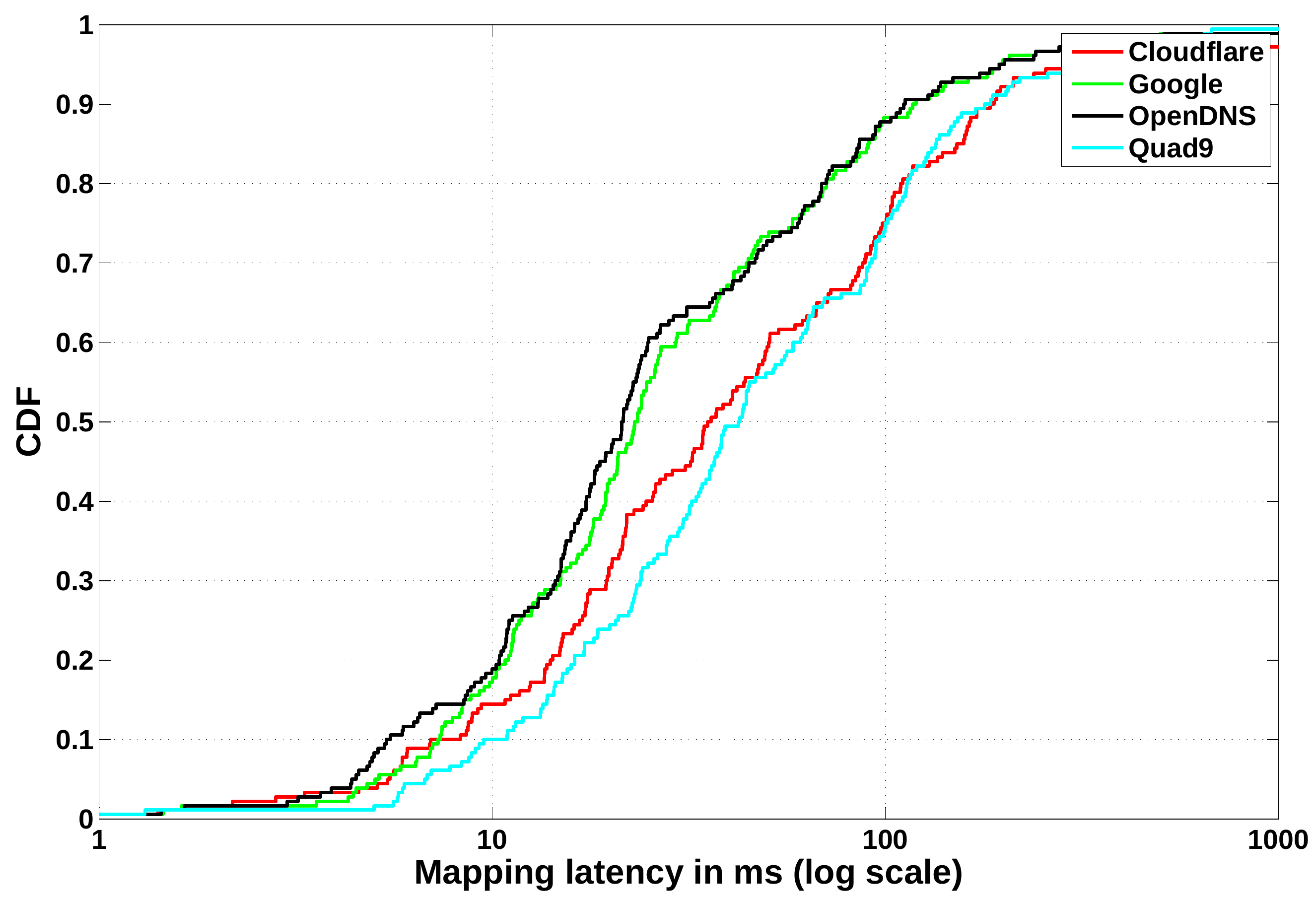}}         
     \end{subfigure}
     \begin{subfigure}[Other CDNs (average per-vantage point latencies)]
         {\includegraphics[width=0.7\columnwidth]{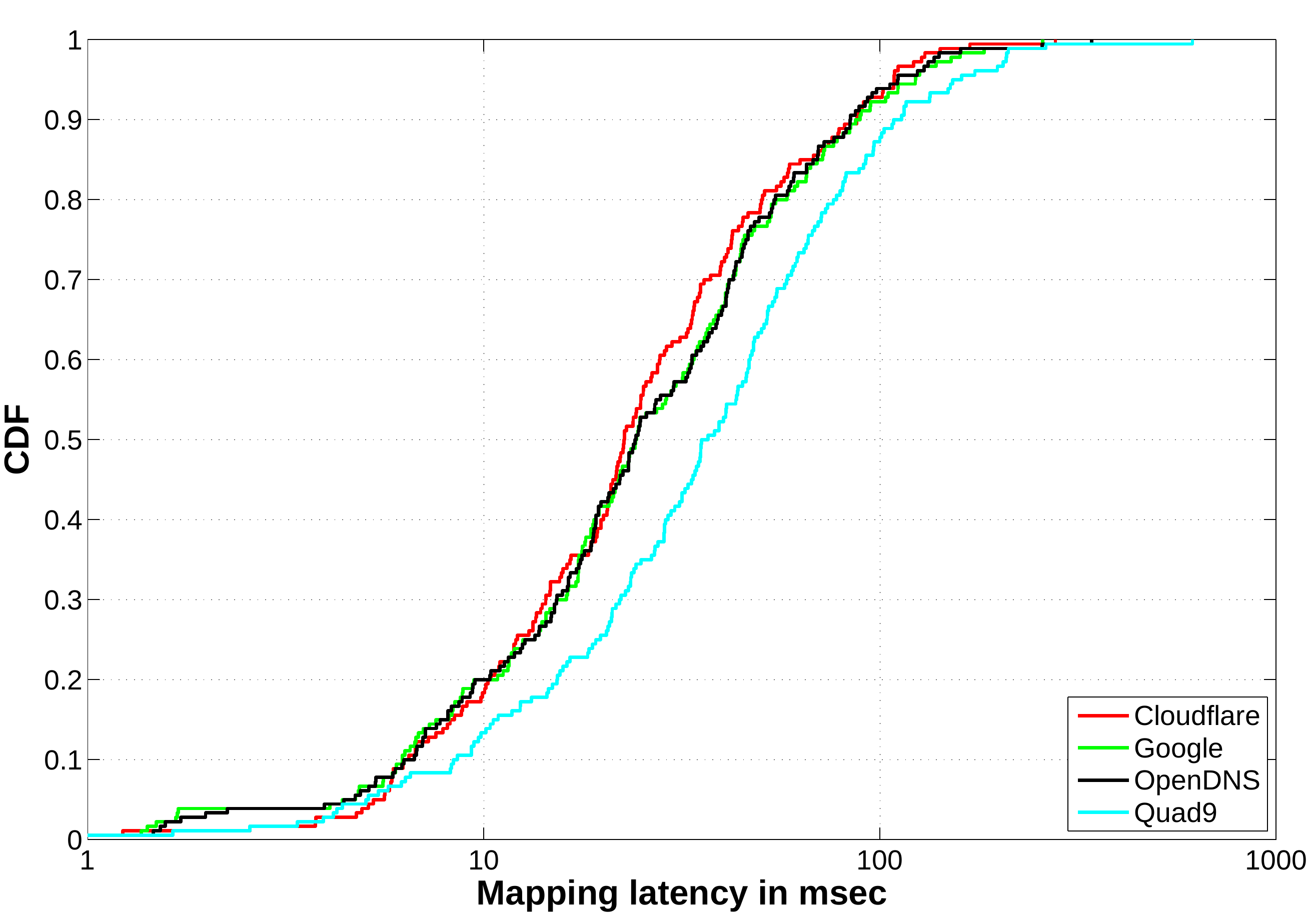}}  
     \end{subfigure}
      \caption{Distribution of IPv4 mapping latencies obtained over IPv4.} 
      \label{fig:mapping_A}
\end{figure*}

\begin{figure*}[tb]
    \centering     
     \begin{subfigure}[Akamai (all latencies)]
         {\includegraphics[width=0.7\columnwidth]{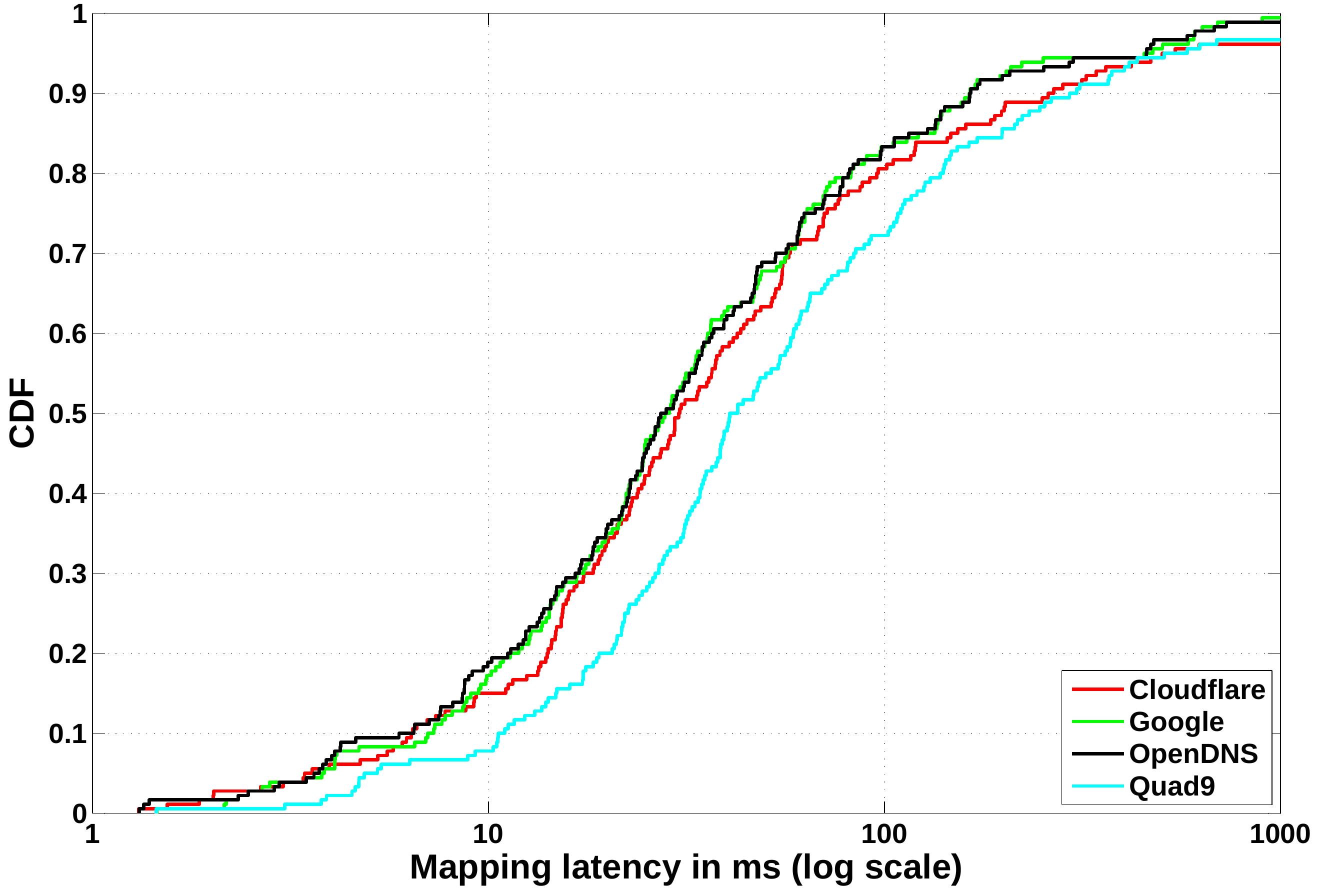}}         
     \end{subfigure}
     \begin{subfigure}[Other CDNs (average per-vantage point latencies)]
         {\includegraphics[width=0.7\columnwidth]{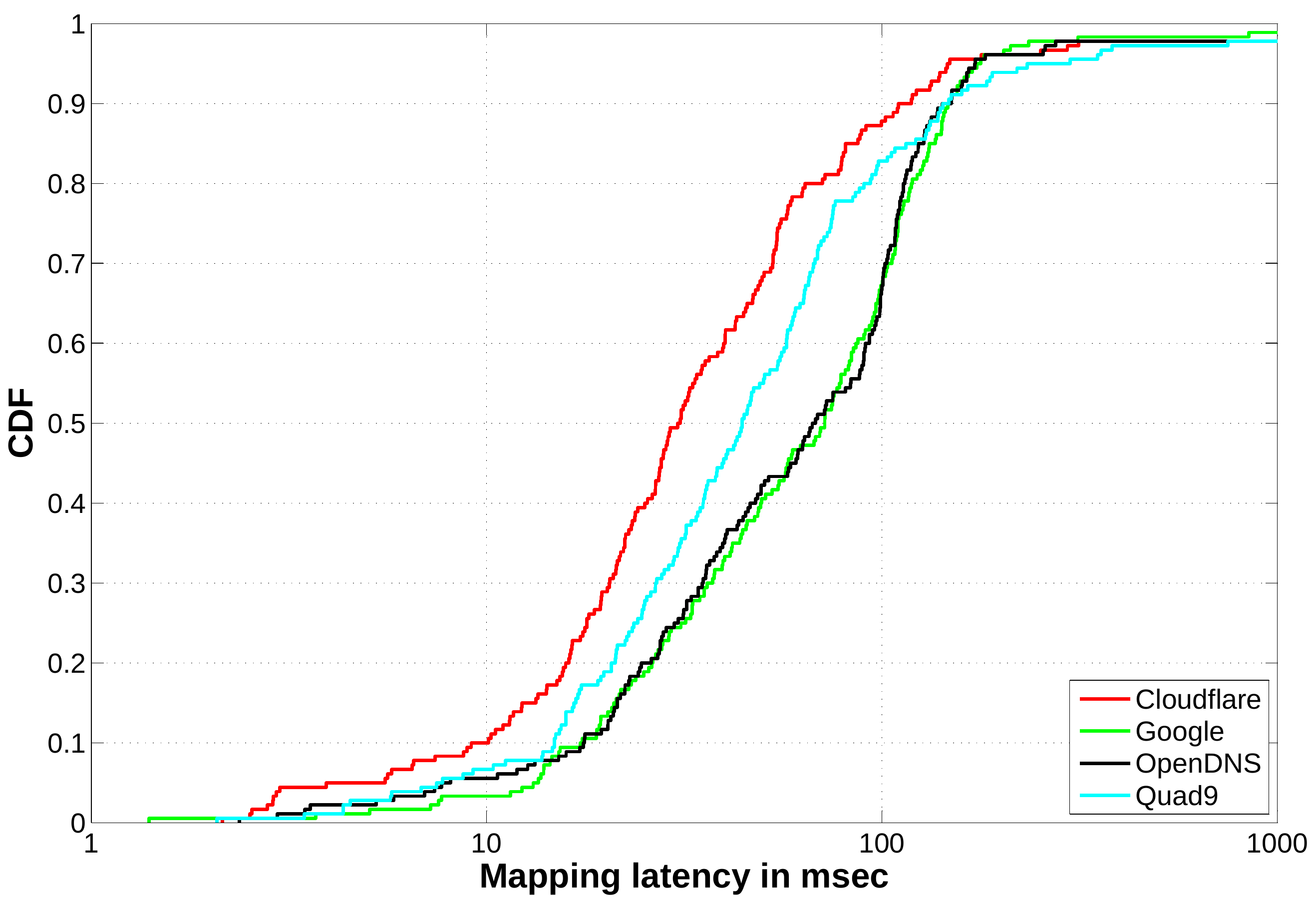}}  
     \end{subfigure}
      \caption{Distribution of IPv6 mapping latencies obtained over IPv6.} \vspace{-2mm}
      \label{fig:mapping_AAAA}
\end{figure*}

\iftoggle{techreport}{
\begin{figure}[t]
    \centering       \includegraphics[width=0.7\columnwidth]{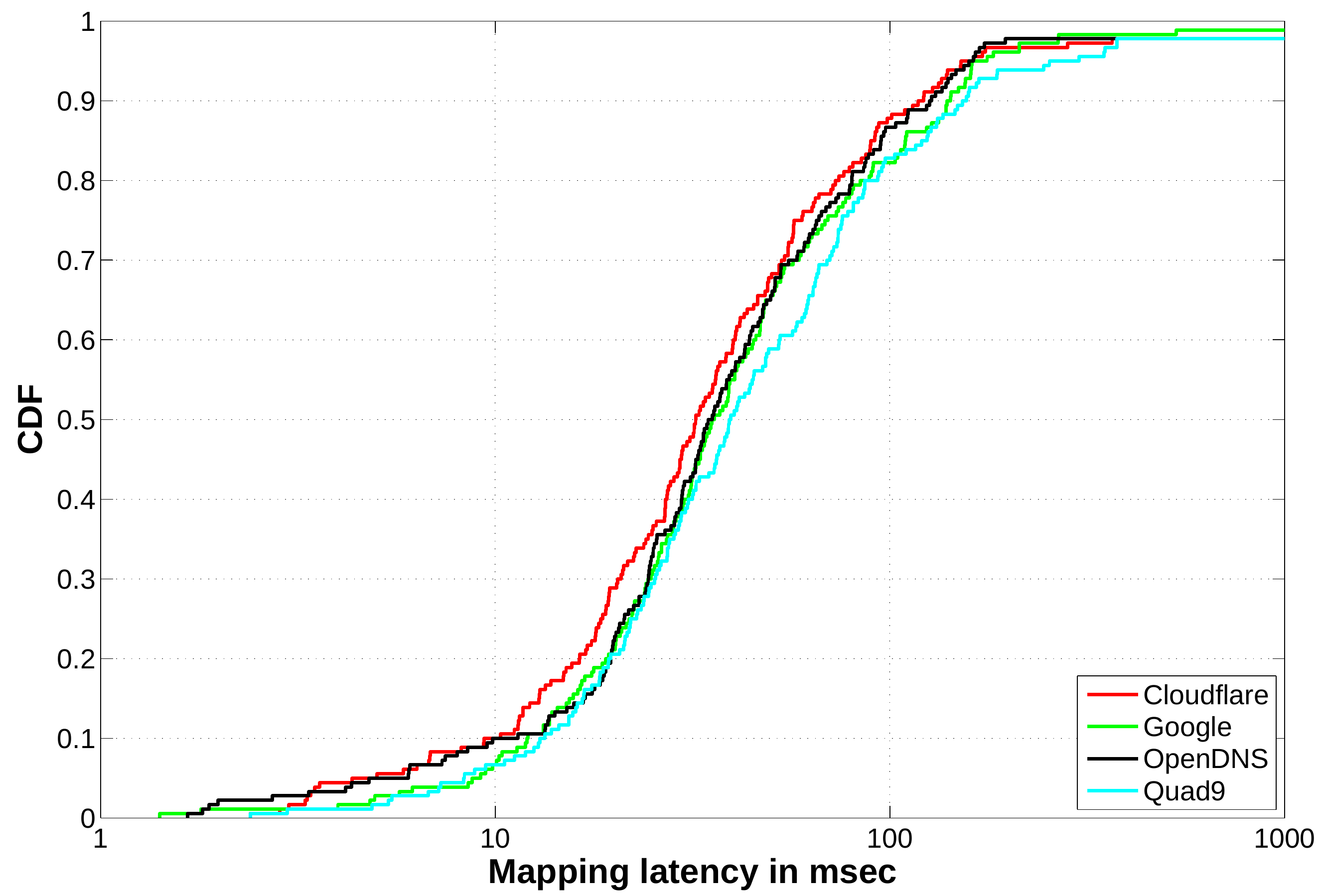}
    \vspace{-1em}
      \caption{Distribution of IPv6 mapping latencies for Cloudfront, Google, and Incapsula CDNs.} \vspace{-5mm}
      \label{fig:mapping_AAAA_all_but_Akamai_and_Fastly}
\end{figure}
}{}

DNS latency represents a fixed cost in the beginning of user interaction with the Web server. The rest of user experience is determined by the quality of CDN's edge server selection.  We now consider how the choice of the resolution services and IP version affects the user-to-edge-server mapping quality produced by the CDNs under study.   Because Akamai's very distinct approach to content delivery\footnote{Akamai uses drastically wider edge server distribution, with over 3,300 locations \cite{maggs_keynote_2019}, while the other CDNs in this study concentrate their platforms in several dozens or low hundreds of locations.}, we consider Akamai separately from the rest of the CDNs.


\subsection{Comparing Public Resolvers}
\label{sec:latency_comparison_publicResolvers}

Figure~\ref{fig:mapping_A} shows CDFs of latencies of IPv4 mappings we obtain by sending an A-type query to resolvers over IPv4, while Figure~\ref{fig:mapping_AAAA} presents the latency distributions of IPv6 mappings obtained over IPv6. For Akamai graphs, data points represent mapping latencies obtained for every Akamai-powered hostname for each vantage point, so each vantage point contributes 65 data points.  The non-Akamai graphs represent distribution of aggregate performance across all non-Akamai CDNs under study.  In these graphs, we would like to prevent CDNs represented by multiple websites from unduly influencing the overall results.  Thus, for each vantage point, we compute the average mapping latency for all websites accelerated through a given CDN, and then plot average latencies for the four CDNs, so that each vantage point contributes one data point. 
The medians of these distributions are listed in Table~\ref{tbl:medianTCPLatencies} in rows for all vantage points. 
We omit distributions of IPv4 mapping latencies obtained over IPv6 and IPv6 mapping latencies over IPv4 as our primary focus on observing any differences between fully IPv4 and IPv6 communication.  

Considering IPv4 communication, Figure~\ref{fig:mapping_A}a shows noticeably lower latencies Akamai mappings obtained through Google and OpenDNS over Cloudflare and Quad9.  We attribute this to the use of ECS by Google and OpenDNS.  For non-Akamai CDNs (Figure~\ref{fig:mapping_A}b), which due to their more concentrated platforms are less sensitive to location mismatch between clients and their resolvers, Cloudflare joins Google and OpenDNS in providing similar mapping quality, while Quad9 still lags behind. We believe  Cloudflare's wider footprint, which allows its egress resolvers to provide better approximation for client locations, is responsible for this finding. This  shows that, with suitable footprint, competitive CDN mappings can be achieved without the ECS extension and related privacy concerns \cite{kintis2016understanding} for all but the widest-distributed CDN platforms.  Of course, from the provider perspective, ECS facilitates good mappings with less extensive footprint and hence at a lower cost.  

Turning to IPv6, Akamai mapping latencies still show lagging Quad9 performance, while the rest of the resolvers produce very close distributions.  In particular, Google and OpenDNS have virtually identical mapping latency distributions despite conveying different ECS prefix lengths (56 for Google and 48 for OpenDNS).   This finding provides a preliminary indication that 48-bit ECS prefix might be sufficient to facilitate client mapping by CDNs, and longer prefixes unnecessarily reduce DNS cache hit rates and leak client information.  A separate in-depth study is needed to reach a more definitive conclusion in this regard.  At the same time, the latency distribution of non-Akamai CDN mappings shows clear separation, with Cloudflare producing the best mappings, followed by Quad9, and then Google and OpenDNS.  This might seem counter-intuitive, as non-Akamai CDNs, with fewer locations, should be less sensitive to the less sensitive to the choice of the resolution services used by a vantage point. We believe the reason for this behavior is that Fastly uses anycast-based, rather than DNS-based, edge server for Google and OpenDNS\footnote{We find that Fastly returns the same four IP addresses to queries from all probes through these resolvers.  An explicit confirmation of anycast use for Google can also be found in the following forum discussion: https://support.fastly.com/hc/en-us/community/posts/360040446972-Fastly-is-suboptimal-when-using-Google-DNS.}.  We verified that the separation disappears once Fastly is removed from the analysis\iftoggle{techreport}{.  Indeed,  Figure~\ref{fig:mapping_AAAA_all_but_Akamai_and_Fastly} shows the same CDFs of IPv6 mapping latencies as in Figure~\ref{fig:mapping_AAAA} but with Fastly removed. All resolvers produce mapping latency distributions that closely track each other.}{ (the graph is in the technical report \cite{full_paper}).}  We note that Fastly has this behavior for IPv4 clients as well, but the effect was not as stark. Our finding that, with sufficiently small distance between clients and their resolvers, anycast produces higher-latency mappings than explicit DNS-based server selection is interesting and deserves a separate study.  

In summary, established resolvers, Google and OpenDNS, produce better user-to-edge-server mappings with Akamai for IPv4 clients than Cloudflare and Quad9, and Cloudflare catches up with Google and OpenDNS for non-Akamai CDNs. The IPv6 mapping quality for Akamai is very similar with all public resolvers except for Quad9, which still lags behind. However, for non-Akamai CDNs and IPv6 clients, Cloudflare and Quad9 show overall better mapping quality, although this is due entirely to Fastly's approach to client request routing. 

\subsection{Public vs. ISP Resolvers}
We compare the quality of CDN mappings obtained through the use of public resolvers and the ISP-provided resolvers.  For fair comparison, we limit this analysis to only the vantage points that we could infer with high confidence used an ISP resolver as their default resolution service.

\iftoggle{techreport}{The "Global" section of Table~\ref{tbl:medianTCPLatencies}, the "w/ISP resolvers" lines, compare median mapping latencies obtained through ISP and public resolvers.}{The "w/ISP resolvers" lines of Table~\ref{tbl:medianTCPLatencies} compare median mapping latencies obtained through ISP and public resolvers.
}  According to the table, public resolvers have largely closed the gap with ISP resolvers when it comes to the CDN IPv4 mapping latency, which was reported in multiple prior studies \cite{ager2010comparing,huang2011public,hours2016study}.  In particular, Google and OpenDNS have completely closed the gap with ISP resolvers on IPv4 mapping quality, as well as IPv6 mapping quality for Akamai.  (They still lag behind in IPv6 mapping for non-Akamai CDNs, which -- as discussed in Section~\ref{sec:latency_comparison_publicResolvers} -- is due to poor performance with Fastly.) And CloudFlare has closed the gap with ISP resolvers on both IPv4 and IPv6 mappings for non-Akamai CDNs, although it is still 5-6 ms (26-31\%) behind, at the median, for Akamai.  Quad9 has similar median IPv6 latencies for non-Akamai CDNs but is further behind in other cases.  While understanding how these remaining gaps translate into quality of experience differences remains a question for future work, 25-50\% higher mapping latency will undoubtedly affect TCP throughput even if the latency difference in absolute values may seem small (5-10 ms at the median).  

\subsection{IPv4 vs. IPv6 Communication}

We now consider the CDN mapping quality of a single-stack client that uses IPv4 or IPv6 throughout its Internet access. In other words, we consider the IPv4 mapping latencies obtained over IPv4 with the IPv6 mapping latencies obtained over IPv6. The corresponding latency distributions can be compared across Figures~\ref{fig:mapping_A}a and \ref{fig:mapping_AAAA}a for Akamai, and ~\ref{fig:mapping_A}b and ~\ref{fig:mapping_AAAA}b for other CDNs.  For clarity, the medians of these distributions are listed in \iftoggle{techreport}{the "Global" section of}{} Table~\ref{tbl:medianTCPLatencies}, the "All" lines.

The table shows that both Akamai and, especially, non-Akamai CDN mappings exhibit clear penalty in mapping latency for IPv6 communication.  Across all vantage points, the Akamai IPv6 penalty is 4-6 ms or between 17--41\% at the median, depending on the resolver, over IPv4.  The other CDNs show similar IPv6 mapping latency penalties for Quad9 and Cloudflair, with  IPv6 mapping latencies around 8 ms or 22-36\% higher at the median than IPv4 latencies. However, the mappings obtained with Google and OpenDNS show dramatically higher IPv6 penalties of 43-47ms, or 1.8--2 times higher than IPv4 latencies at the median. 
These penalties are drastically higher those than noted by Bajpai et al. \cite{bajpai2016measuring} for general websites, who observed that over 90\% of TCP latencies to these websites over IPv6 were within 1ms of those over IPv4. We leave understanding the reason for these high IPv6 penalties for future work.  

Finally we note that the IPv6 penalties, while sometimes significant, in most cases still stay below the thresholds in the Happy Eyeballs protocol, which falls to IPv4 only if IPv6 TCP latency is higher by a threshold with a recommended default value of 250 ms \cite{rfc8305}. Thus, the IPv6 mapping penalty we found would not lead a dual-stack client to switch to IPv4, relegating the browser to use higher-latency mapping for the duration of the HTTP interaction.  



\iftoggle{techreport}{
\begin{table*}[t]
\footnotesize
 \begin{center}
\scalebox{0.9}{
 \begin{tabular}{|c|c|c|c|c|c|c|c|c|c|c|c|}
  \hline
Region&IP   & \multicolumn{2}{c|}{ISP-provided} & \multicolumn{2}{c|}{Cloudflare} & \multicolumn{2}{c|}{Quad9}  & \multicolumn{2}{c|}{Google} & \multicolumn{2}{c|}{OpenDNS} \\ \cline{3-12}
& & A & AAAA & A & AAAA & A & AAAA & A & AAAA & A & AAAA \\ \hline \hline
N. America &v4 & 2.57& 2.43 & 9.15 & 9.27 &13.44 & 11.58 & 37.98 &39.92 & 33.98 & 32.72 \\ \cline{2-12}
     & v6& 3.08& 3.06 & 9.49& 9.50 & 10.63 & 10.91 & 41.52 & 40.57 & 29.84 & 28.59\\\hline  \hline
Europe &v4 & 2.24& 2.21 & 15.23 & 15.22 & 38.63 & 37.74 & 50.39 & 52.30 & 52.33 & 50.49 \\ \cline{2-12}
     & v6& 2.26 &2.28 & 18.09& 18.16 & 36.59 & 36.56 &52.65 &52.10 & 51.30 & 49.69\\\hline \hline
Latin America &v4 & 13.52& 13.45 & 32.08 &32.26 & 152.78 & 152.48 & 96.79 &100.06 & 195.73 & 192.16 \\ \cline{2-12}
     & v6& 13.22 &13.22 &63.75& 63.95 & 152.57 & 153.13 & 104.77 & 104.24 & 164.19 & 153.63\\\hline \hline
Asia &v4 & 3.12& 3.11 &27.73 & 27.65 &81.67 & 80.92 &86.56 &89.88 & 108.25 & 103.97 \\ \cline{2-12}
     & v6& 2.23 & 2.35 & 45.82& 45.66 & 87.43 &88.24 & 86.93 &86.48 & 101.56& 99.68\\\hline  \hline
Africa &v4 & 1.89& 1.88 & 49.05 & 49.02 & 4.85 & 4.38 & 144.60 & 145.53 & 231.70 & 224.82 \\ \cline{2-12}
     & v6& 4.09& 3.87 & 7.65& 7.66 & 4.37 & 4.67 & 144.85 & 144.11 & 217.70 & 213.29\\\hline
Oceania &v4 & 2.62& 2.28 &13.08 &13.12 & 12.89 & 12.63 & 157.73 & 163.52 & 168.96 & 146.20 \\ \cline{2-12}
     & v6& 39.37 & 39.33 & 20.07& 20.00 & 12.46 & 12.41 & 168.48 & 168.48 & 151.31 & 132.22\\\hline   \hline
\end{tabular}
}
  \caption{Median DNS latencies by region (msec).}
  \label{tbl:medianDNSLatencies_cont}
  \end{center}
\end{table*}
}{}

\iftoggle{techreport}{\begin{table*}[t]
\small
 \begin{center}
 \begin{tabular}{|c|c|c|c|c|c|c|c|c|c|c|c|c|}
  \hline
Region & {CDN} & Vantage  & \multicolumn{2}{c|}{ISP-provided} & \multicolumn{2}{c|}{Cloudflare} & \multicolumn{2}{c|}{Quad9}  & \multicolumn{2}{c|}{Google} & \multicolumn{2}{c|}{OpenDNS} \\ \cline{4-13}
  & & points & IPv4    & IPv6  & IPv4     & IPv6  & IPv4      & IPv6 & IPv4     & IPv6 &  IPv4 & IPv6 \\ \hline \hline
 & Akamai & All  &      &       & 18.93    & 23.22    & 29.01 & 34.09    & 15.08    & 20.83   & 14.32  & 20.23 \\ \cline{3-13}
 Global      &         &  w/ISP resolver &  18.23    &  18.71     & 23.02    & 24.47    & 27.22 & 28.49    & 18.58    & 18.94   & 17.93  & 18.71 \\ \cline{2-13}
    & Others & All   &       &      & 22.59    & 30.66    & 36.12 & 44.27    & 24.19    & 71.66    & 24.07  & 67.27  \\ \cline{3-13}
    &  &  w/ISP resolver   &  27.08     & 34.85     & 25.09    & 35.30    & 34.16 & 34.93    & 28.01    & 76.66    & 24.81  & 96.02  \\ \hline \hline
 & Akamai & All  &      &       & 14.99    & 15.93    & 18.00 & 17.95    & 12.49    & 18.36   & 12.76  & 18.34 \\ \cline{3-13}
North       &         &  w/ISP resolver &  19.02    &  18.50     & 22.47    & 28.67    & 22.51 & 29.58    & 19.22    & 18.99   & 19.47  & 19.00 \\ \cline{2-13}
America   & Others & All   &       &      & 16.61    & 20.93    & 18.29 & 24.71    & 17.33    & 37.61    & 16.35  & 34.28  \\ \cline{3-13}
  &  &  w/ISP resolver   &  26.89     & 34.66     & 21.57    & 29.00    & 22.79 & 31.95    & 24.77    & 35.52    & 21.23  & 35.82  \\ \hline \hline
 & Akamai & All  &      &       & 16.70    & 18.70    & 25.08 & 31.57    & 14.08    & 18.08   & 13.48  & 17.87 \\ \cline{3-13}
 Europe      &         &  w/ISP resolver &  14.06    &  13.86     & 19.04    & 21.59    & 22.92 & 21.57    & 14.87    & 17.03   & 14.44  & 13.88 \\ \cline{2-13}
    & Others & All   &       &      & 21.15    & 25.53    & 40.27 & 42.14    & 21.89    & 45.58    & 21.56  & 40.57  \\ \cline{3-13}
    &  &  w/ISP resolver   &  22.47     & 25.23     & 18.66    & 23.69    & 27.06 & 26.96    & 21.88    & 97.69    & 21.54  & 92.94  \\ \hline \hline
 & Akamai & All  &      &       & 34.22    & 48.00    & 151.39 & 158.21    & 27.56    & 41.57   & 22.11  & 41.61 \\ \cline{3-13}
 L.America      &         &  w/ISP resolver &  15.87    &  105.32     & 81.95    & 113.68    & 186.36 & 239.69    & 22.51    & 106.51   & 17.71  & 106.58 \\ \cline{2-13}
    & Others & All   &       &      & 70.07    & 63.91    & 148.60 & 140.64    & 89.63    & 94.39    & 91.81  & 88.01  \\ \cline{3-13}
    &  &  w/ISP resolver   &  278.39     & 147.82     & 94.28    & 119.29    & 208.93 & 351.39    & 256.96    & 144.72    & 256.71  & 110.85  \\ \hline \hline
 & Akamai & All  &      &       & 31.03    & 54.47    & 65.41 & 66.35    & 23.09    & 38.79   & 20.77  & 39.60 \\ \cline{3-13}
 Asia     &         &  w/ISP resolver &  38.81    &  40.67     & 34.65    & 54.55    & 45.55 & 45.75    & 41.43    & 37.23   & 38.67  & 36.82 \\ \cline{2-13}
    & Others & All   &       &      & 35.27    & 59.19    & 67.49 & 84.35    & 43.36    & 108.26    & 45.43  & 108.75  \\ \cline{3-13}
    &  &  w/ISP resolver   &  45.95     & 78.40     & 35.27    & 42.90    & 79.39 & 60.14    & 43.36    & 78.08    & 43.13  & 108.75  \\ \hline \hline
 & Akamai & All  &      &       & 52.20    & 49.12    & 51.81 & 49.36    & 52.25    & 49.15   & 49.63  & 49.14 \\ \cline{3-13}
 Africa      &         &  w/ISP resolver &  52.34    &  18.83     & 116.24    & 16.61    & 52.23 & 52.19    & 52.54    & 18.78   & 52.26  & 18.81 \\ \cline{2-13}
    & Others & All   &       &      & 121.64    & 118.77    & 108.20 & 132.50    & 77.98    & 165.49    & 82.17  & 158.66  \\ \cline{3-13}
    &  &  w/ISP resolver   &  83.47     & 76.35     & 129.81    & 133.44    & 116.38 & 142.72    & 126.05    & 173.98    & 129.41  & 158.66  \\ \hline \hline
 & Akamai & All  &      &       & 15.92    & 15.27    & 20.35 & 14.64    & 14.78    & 14.10   & 14.09  & 14.10 \\ \cline{3-13}
 Oceania     &         &  w/ISP resolver &  12.98    &  38.54     & 41.30    & 13.24    & 38.61 & 38.55    & 13.84    & 13.51   & 13.34  & 13.45 \\ \cline{2-13}
    & Others & All   &       &      & 34.06    & 57.36    & 39.25 & 57.54    & 43.17    & 154.87    & 46.45  & 100.63  \\ \cline{3-13}
    &  &  w/ISP resolver   &  38.19     & 58.08     & 34.06    & 87.01    & 73.68 & 63.71    & 40.76    & 203.53    & 46.45  & 166.01  \\ \hline \hline
    \end{tabular}
  \caption{Median mapping latencies (ms).  The IPv4 (resp. IPv6) columns show the latencies of IPv4 (resp. IPv6) mappings obtained when using IPv4 (resp. IPv6) to interact with resolvers.} \vspace{-6mm}
  \label{tbl:medianTCPLatencies}
  \end{center}
\end{table*}
}{\begin{table*}[t]
\small
 \begin{center}
 \begin{tabular}{|c|c|c|c|c|c|c|c|c|c|c|c|}
  \hline
 {CDN} & Vantage  & \multicolumn{2}{c|}{ISP-provided} & \multicolumn{2}{c|}{Cloudflare} & \multicolumn{2}{c|}{Quad9}  & \multicolumn{2}{c|}{Google} & \multicolumn{2}{c|}{OpenDNS} \\ \cline{3-12}
& points & IPv4    & IPv6  & IPv4     & IPv6  & IPv4      & IPv6 & IPv4     & IPv6 &  IPv4 & IPv6 \\ \hline \hline
Akamai & All  &      &       & 18.93    & 23.22    & 29.01 & 34.09    & 15.08    & 20.83   & 14.32  & 20.23 \\ \cline{2-12}
         &  w/ISP resolver &  18.23    &  18.71     & 23.02    & 24.47    & 27.22 & 28.49    & 18.58    & 18.94   & 17.93  & 18.71 \\ \hline
Others & All   &       &      & 22.59    & 30.66    & 36.12 & 44.27    & 24.19    & 71.66    & 24.07  & 67.27  \\ \cline{2-12}
&  w/ISP resolver   &  27.08     & 34.85     & 25.09    & 35.30    & 34.16 & 34.93    & 28.01    & 76.66    & 24.81  & 96.02  \\ \hline 
    \end{tabular} 
    \caption{Median mapping latencies (ms).  The IPv4 (resp. IPv6) columns show the latencies of IPv4 (resp. IPv6) mappings obtained when using IPv4 (resp. IPv6) to interact with resolvers.} \vspace{-6mm}
  \label{tbl:medianTCPLatencies}
  \end{center}
\end{table*}
}

\section{Regional Differences}

We consider how our findings may be affected by regional differences.  
\iftoggle{techreport}{Table~\ref{tbl:medianDNSLatencies_cont} shows the median DNS latencies for each region, while Table~\ref{tbl:medianTCPLatencies} lists the median  TCP latencies of the regional mappings obtained using each service.}{We only present the summary of key differences from the global results and refer to the technical report~\cite{full_paper} for details.} We emphasize again that our finding represent point observations at specific vantage points and should not be viewed as representing general user experiences in these regions, especially in those regions where we had only few vantage points suitable for our measurements, such as in Africa and Latin America.  
One key observation is that developing regions amplify the differences among DNS resolution providers, presumably reflecting uneven infrastructure build-out.  For example, in our Oceania and African locations, Google and OpenDNS have median DNS latencies of between 132 and 231 ms, depending on the protocol and query type, while Quad9's median DNS latencies are under 13 and 5 ms for Oceania and African locations, respectively.  At the same time, across all our Latin American locations, Google and OpenDNS produce Akamai mappings with median latencies of under 42 ms in both IP versions, while Quad9's median mapping latencies are over 150 ms in both versions.  Thus, the choice of resolvers can have a profound performance impact on users outside Europe and North America.

Another interesting finding is that IPv6 penalty in CDN mapping latency, which we observed for all resolvers globally, is absent in our African locations when it comes to Akamai (Google and OpenDNS still have high IPv6 penalty of, resp., 87 and 76 ms or roughly two times higher than IPv4 mappings for non-Akamai CDNs).  Moreover, Cloudflare's DNS latency in our African locations is starkly lower over IPv6 than over IPv4, with median DNS latencies of 49ms and under 8 ms, respectively.  We have no substantiated explanation for this but can speculate that while IPv6 deployment in Africa is generally lagging \cite{GoogleIPv6_adoption}, in the few locations where IPv6 is available, it is implemented using more modern higher performance components than legacy IPv4.

\section{Conclusion}
\label{sec:concl}

This paper studies the performance of DNS resolution  services in the face of significant recent developments on the Internet,  namely,  the  IPv6  finally  getting  traction  and the adoption of the ECS extension to DNS by major DNS and CDN services.  In particular, we consider the performance of the DNS system from the end-user perspective, both in terms of its response latency and in terms of the quality of the DNS-driven mappings of users to CDN edge servers. We find that DNS resolution services differ drastically – by an order of magnitude in some locations – in both of these metrics. We also find established resolvers (Google DNS and OpenDNS) to lag far behind relative newcomers (Cloudflair and Quad9) in terms of DNS latency, while Google and OpenDNS produce better user mappings for Akamai than Cloudflair and Quad9. We further find that, although individual resolvers's performance varies across regions, CDNs, and IP versionspublic resolvers have by and large closed the gap with ISP resolvers in the quality of CDNs’client-to-edge-server mappings as measured by latency.  Finally, in most locations, we observe IPv6 penalty in the latency of client-to-CDN-edge-server mappings produced by the resolvers. Moreover, this penalty, while often significant, still does not rise above typical thresholds employed by the Happy Eyeballs algorithm for preferring IPv4 communication. Thus, dual-stacked clients in these locations may experience suboptimal performance.

Overall, we conclude that clients outside Africa who choose IPv6 for Internet communication, or use the Happy Eyeballs algorithm \cite{rfc8305} to dynamically select between IPv4 and IPv6 for TCP communication, may experience suboptimal performance when accessing CDN-accelerated Web content.  Also, because different DNS resolution service providers can have large effect on DNS performance as well as CDN mappings in different regions, clients should evaluate their providers carefully based on  their specific Internet location.  Providing a convenient tool for users to compare different DNS resolution services is on our plate for future work.

\bibliographystyle{abbrv}
\bibliography{references}
\end{document}